\documentclass[12pt]{article}
\usepackage{graphicx}
\usepackage{subfigure}
\input epsf.tex

\newcommand{\beq}{\begin{equation}}
\newcommand{\eeq}{\end{equation}}
\newcommand{\be}{\begin{equation}}
\newcommand{\ee}{\end{equation}}
\newcommand{\bea}{\begin{eqnarray}}
\newcommand{\eea}{\end{eqnarray}}
\newcommand{\req}[1]{~(\ref{#1})}

\newcommand{\gs}{\mbox{$g_s$}}            
\newcommand{\ap}{\mbox{$\alpha^\prime$}}  
\newcommand{\ls}{\mbox{$l_s$}}            

\def\href#1#2{#2}

\def\p{\partial}

\begin{document}

\baselineskip=15.5pt
\pagestyle{plain}
\setcounter{page}{1}

\begin{titlepage}
\begin{flushleft}
       \hfill                      {\tt hep-th/0809.****}\\
       \hfill                       FIT HE - 08-03 \\
       \hfill                       KYUSHU-HET ** \\
       \hfill                       Kagoshima HE - 08-2 \\
\end{flushleft}
\vspace*{3mm}

\begin{center}
  {\huge Baryonium in Confining Gauge Theories}
\end{center}
\vspace{5mm}

\begin{center}
\vspace*{5mm}
\vspace*{2mm}
\vspace*{5mm}
{\large Kazuo Ghoroku${}^{\dagger}$\footnote[1]{\tt gouroku@dontaku.fit.ac.jp},
Masafumi Ishihara${}^{\ddagger}$\footnote[2]{\tt masafumi@higgs.phys.kyushu-u.ac.jp},
${}^{\S}$Akihiro Nakamura\footnote[3]{\tt nakamura@sci.kagoshima-u.ac.jp}
${}^{\P}$Fumihiko Toyoda\footnote[4]{\tt ftoyoda@fuk.kindai.ac.jp}
%
}\\

{${}^{\dagger}$Fukuoka Institute of Technology, Wajiro, 
Higashi-ku} \\
{
Fukuoka 811-0295, Japan\\}
{
${}^{\ddagger}$Department of Physics, Kyushu University, Hakozaki,
Higashi-ku}\\
{
Fukuoka 812-8581, Japan\\}
{
${}^{\S}$Department of Physics, Kagoshima University, Korimoto 1-21-35, \\Kagoshima 890-0065, Japan\\}
{
${}^{\P}$School of Humanity-Oriented Science and
Engineering, Kinki University,\\ Iizuka 820-8555, Japan}

\vspace*{10mm}
\end{center}

\begin{center}
{\large Abstract}
\end{center}
We show a new class of embedding solutions of D5 brane, which wraps on $S^5$ in the AdS${}_5\times S^5$ space-time and contains fundamental strings as $U(1)$ flux to form a baryon vertex. The new solution given here is different from the baryon vertex since it consists of two same side (north or south) poles of $S^5$ as cusps, which are put 
on different points in our three dimensional space. This implies that the same magnitude of electric displacement exists at each cusp, but their orientations are opposite due to the flux number conservation. This configuration is therefore regarded as a D5-$\overline{D5}$ bound state, and we propose this as the
vertex of a baryonium state, which is made of a baryon and an anti-baryon. By attaching quarks and anti-quarks to the two cusps of this vertex, it is possible to construct a realistic baryonium.

\noindent

\vfill
\begin{flushleft}

\end{flushleft}
\end{titlepage}
\newpage

\section{Introduction}

In the context of string/gauge theory correspondence
\cite{jthroat,gkpads,wittenholo}, 
the baryon has been studied as a system of fundamental strings (F-strings) and 
D5-branes wrapped on $S^5$ in AdS${}_5\times S^5$ space-time \cite{wittenbaryon, 
groguri,imamura,cgs,GRST,cgst,ima-04,baryonsugra,imafirst}. 
They correspond to quarks and the baryon vertex respectively.
The F-strings are partially dissolved as a $U(1)$ flux in the D5 brane, and 
their remaining parts
flow out from one (or two) cusp(s) 
on the surface of the D5 brane as separated free strings. 
The baryon vertex has complicated structures which are
given as solutions of the equations of motion for the D5 brane embedded
in an appropriate background, which is dual to the confining gauge theory
(for example \cite{Liu:1999fc,KS2,GY}). 
This picture has been recently studied furthermore
\cite{GI} along the Born-Infeld approach given in \cite{imamura}-\cite{cgst},
and also extended to finite temperature theory \cite{GINT,ALR}

Here, we show new kinds of configurations, which are obtained as solutions 
of the same equations with the one which gives
baryon vertex solutions. But the new solutions given here are different from the
baryon vertex. The baryon vertex wraps whole $S^5$ once. Namely, it
covers all range of the polar angle ($\theta$) of the 
$S^5$, $0\leq \theta\leq\pi$, once.

On the other hand, the configuration of the new solution 
covers twice one polar side, 
for example in the range of $\theta_0\leq \theta\leq\pi$ where $0<\theta_0$. And,
at any $\theta$ in this range, this configuration 
exists at two different point in our three dimensional space. Then 
this configuration looks like a line with a finite length. On this line,
the point of $\theta=\theta_0$ is at its center and the two end points
are given by the same $\theta=\pi$. In other words, it starts from one end 
at $\theta=\pi$ and arrives 
at $\theta=\theta_0$ (the center of the configuration) then goes
to the other end point $\theta=\pi$ along another half path.
This solution can be interpreted as the connection of two $U(1)$ fluxes with 
opposite charge. 
This fact implies that it describes the same polar sides ($\theta=\pi$) of
connected two $S^5$s wrapped by D5 and anti-D5 ($\overline{D5}$) branes
respectively. 

This can be regarded as a $D5/\overline{D5}$ bound states.
Similar D-brane embeddings
have been found for different D-branes in different backgrounds
\cite{KMMW2,SS,SSu}. 
The essential point of such solutions is that the configuration covers two different
points in our space at the same point of
a world volume coordinate (here $\theta$) of the D-brane. 
In this sense, our solution
is essentially the same type with the former examples.

The two end points are the cusps, where opposite sign of $U(1)$ fluxes exist.
Then the F-strings attached at these two cusps have also
opposite orientations with the same number. This is considered
as the baryonium or the bound state of a baryon and an anti-baryon. 

The energy and the configuration
of this baryonium vertex depend on boundary conditions of the equations
of motion. So, varying the boundary conditions, the relation
between the vertex energy and the distance of the two cusps is examined. 
And we could find a minimum vertex energy at a finite distance between the two cusps.
This implies that the baryon and anti-baryon bound state is
stable against vanishing to the vacuum. 

In Section \ref{eqnsec} we give our model and D5-brane action
with non-trivial $U(1)$ gauge field. 
And the equations of motion for D5 branes are given. 
In section
\ref{bound}, we give $D5/\overline{D5}$ solutions as baryonium
vertex. And its configuration and energy, which depends on the configurations,
are examined. In the section 4, the differences between the baryonium and 
split baryons are discussed. 
And 
in the final section, we summarize our results and discuss related
directions.

\section{Model}\label{eqnsec}

\subsection{Bulk background}

We derive $D5$-$\overline{D5}$ solutions as baryonium from the equations 
of motion given by the action of D5-brane which 
is embedded in a supersymmetric
10d background of type IIB theroy. 
The background solution should be dual to the confining gauge theory since
the baryonium examined here is a bound state of quarks. 
While there may be some
such solutions,
we consider the following background \cite{Liu:1999fc,KS2,GY},
\beq
ds^2_{10}= e^{\Phi/2}
\left(
\frac{r^2}{R^2}\eta_{\mu\nu}dx^\mu dx^\nu +
\frac{R^2}{r^2} dr^2+R^2 d\Omega_5^2 \right) \ ,
\label{non-SUSY-sol}
\eeq
which is written in string frame. At the same time, the dilaton $\Phi$ and
the axion $\chi$ are given as
\beq 
 e^\Phi= 
1+\frac{q}{r^4} \ , \quad \chi=-e^{-\Phi}+\chi_0 \ ,
\label{dilaton} 
\eeq
and with self-dual Ramond-Ramond field strength
\begin{equation}\label{fiveform}
G_{(5)}\equiv dC_{(4)}=4R^4\left(\mbox{vol}(S^5) d\theta_1\wedge\ldots\wedge d\theta_5
-{r^3\over R^8}  dt\wedge \ldots\wedge dx_3\wedge dr\right),
\end{equation}
where $\mbox{vol}(S^5)\equiv\sin^4\theta_1
\mbox{vol}(S^4)\equiv\sin^4\theta_1\sin^3\theta_2\sin^2\theta_3\sin\theta_4$.

This solution, (\ref{non-SUSY-sol})-(\ref{dilaton}),
is useful since the confinement
of quarks are realized due to the gauge condensate $
q\equiv \langle F_{\mu\nu}^2\rangle$ \cite{KS2,GY}, which is given by the coefficient
of $1/r^4$ for the asymptotic expansion of the dilaton at large $r$.
And furthermore, $\cal{N}$=2 supersymmetry is preseved in spite of the non-trivial
dilaton is introduced.
We can assure through the Wilson loop that $q^{1/2}$ is proportional to
the tention of the linear rising potential between the quark 
and anti-quark \cite{GY}. In the present case, $q$ is essential to fix
the size of the baryonium and stabilize it energetically as shwon below. 
 
We notice that
the axion $\chi$ corresponds to the souce of D(-1) brane and it is Wick rotated
in the supergravity action. This is necessary to preserve the supersymmetry.

\subsection{$D5$ brane action}
\label{eqnsec2}
The baryon is constructed from the vertex and $N$ fundamental strings, and
the vertex is given by the D5 brane wrapped on the
${S}^{5}$ of the above metric.
The $N$ fundamental strings terminate on this vertex and they are dissolved
in it \cite{wittenbaryon,groguri} as $U(1)$ flux. 
The D5-brane action is thus written as by the Dirac-Born-Infeld (DBI) plus
WZW term \cite{cgs}
\begin{eqnarray}\label{d5action}
S_{D5}&=&-T_{5}\int d^6\xi
 e^{-\Phi}\sqrt{-\det\left(g_{ab}+\tilde{F}_{ab}\right)}+T_{5}\int
d^6\xi \tilde{A}_{(1)}\wedge C_{(5)} ~,\\
g_{ab}&\equiv&\p_a X^{\mu}\p_b X^{\nu}G_{\mu\nu}~, \qquad
C_{a_1\ldots
a_5}\,\equiv\,\p_{a_1}X^{\mu_1}\ldots\p_{a_5}X^{\mu_5}G_{\mu_1\ldots\mu_5}~.\nonumber
\end{eqnarray}
where $\tilde{F}_{ab}=2\pi\ap F_{ab}$ and 
$T_5=1/(\gs(2\pi)^{5}\ls^{6})$ is the brane tension.

\vspace{.3cm}
The D5 brane is embedded in the world volume
$\xi^{a}=(t,\theta,\theta_2,\ldots,\theta_5)$, where $(\theta_2,\ldots,\theta_5)$
are the $S^4$ part with the volume of $\Omega_{4}=8\pi^{2}/3$, where we set
as $\theta_1=\theta$. 
Restrict our attention to $SO(5)$ symmetric configurations of the 
form $r(\theta)$, $x(\theta)$, and $A_t(\theta)$ (with all other fields 
set to zero). 
Then the above action is written as
\be \label{d3action}
S= T_5 \Omega_{4}R^4\int dt\,d\theta \{ -\sin^4\theta 
  \sqrt{e^{\Phi}\left(r^2+r^{\prime 2}+(r/R)^{4}x^{\prime 2}\right)
   -\tilde{F}_{t \theta}^2}  -\tilde{F}_{t \theta} D \},
\ee
where the WZW term is rewritten by partial integration with respect to
$\theta$, and $\Omega_{4}=8\pi^{2}/3$ is the volume of the unit four-sphere.
The factor $D(\theta)$ is defined by
\beq\label{D}
\partial_\theta D = -4 \sin^4\theta~,
\eeq
and is related to $\tilde{F}_{t \theta}$ by the equation of motion for
$\tilde{A}_{t}$ as
\beq\label{D2}
  D={\sin^4\theta~ \tilde{F}_{t \theta}\over
  \sqrt{e^{\Phi}\left(r^2+r^{\prime 2}+(r/R)^{4}x^{\prime 2}\right)
   -\tilde{F}_{t \theta}^2}}~.
\eeq
We call this $D$ as displacement, and it is given by 
solving (\ref{D}) as follows ,
\be \label{d}
  D(\nu,\theta) \equiv \left[{3\over 2}(\nu\pi-\theta)
  +{3\over 2}\sin\theta\cos\theta+\sin^{3}\theta\cos\theta\right].
\ee
The meaning of the integration constant, defined in the
range of $0\leq\nu\leq 1$, is given below.

Next, the action is rewritten by eliminating the gauge field
in terms of (\ref{D2}) to obtain an energy
functional of the embedding coordinate only 
\footnote{
$U$ is obtained by a Legendre transformation of $L$, which is defined
as $S=\int dt L$, as $U={\partial L\over \partial\tilde{F}_{t \theta}}
\tilde{F}_{t \theta}-L$. Then equations of motion of (\ref{d3action}) 
provides the same solutions of the one of $U$. 
}:
\be \label{u}
U = {N\over 3\pi^2\alpha'}\int d\theta~e^{\Phi/2}
\sqrt{r^2+r^{\prime 2} +(r/R)^{4}x^{\prime 2}}\,
\sqrt{V_{\nu}(\theta)}~.
\ee
\be\label{PotentialV}
V_{\nu}(\theta)=D(\nu,\theta)^2+\sin^8\theta
\ee
where we used $T_5 \Omega_{4}R^4=N/(3\pi^2\alpha')$.
Using this expression (\ref{u}), we consider the meaning of the 
integration constant $\nu$ given in (\ref{D}). In the below, we solve the 
equation of motion for $r(\theta)$ and we find that it has two cusps
or singular points at $r(\theta)=r(\pi)$ and $r(0)$, namely at $\theta=\pi$ and
$\theta=0$. At these points, $r'=\partial_{\theta}r$ diverges 
and $x'\simeq 0$ for $q=0$. The configuration near these positions 
represents the bundle of the fundamental strings. The numbers of the 
fundamental strings at the cusps are estimated as follows. At $\theta=\pi$
and for $q=0$ ($\Phi=0$), we obtain the following approximate formula
\be \label{upi}
U \simeq {N\over 3\pi^2\alpha'}\int dr~{3\over 2}(1-\nu)\pi~
={N\over 2\pi\alpha'}(1-\nu)\int dr~.
\ee
And similary, we obtain the following at $\theta=0$,
\be \label{0}
U \simeq {N\over 2\pi\alpha'}\nu\int dr~.
\ee
Since ${1\over 2\pi\alpha'}\int dr$ represents the bundle of a fundamental 
string, the total number of fundamental strings is given by $N$, which
are separated to $N(1-\nu)$ and $N\nu$ to each cusp point. The meaning of
$\nu$ is then the ratio of this separation, so it must be defined as 
$0\leq\nu(\equiv k/N)\leq 1$, where $k(\leq N)$ is an integer.

{By the definition of $U$, Eq.(\ref{u}), $U$ is positive, and it is proportional
to $|D(\nu,0)|$ or $|D(\nu,\pi)|$.
Then the total number
of the flux is counted as $N$ when we sum up the one of the two cusps at $\theta=0$
and $\theta=\pi$. However, we notice here the orientation of the flux of $U(1)$
current, then $D$ defined by (\ref{D2}) could takes two possible value,
$D=\pm |D|$, depending on the orientation of the flux. For the case of opposite
orientation, the total flux number would be counted as $-N$. This is regarded
as the anti-baryon vertex.

Then two possible flux numbers are assinged as $\pm N(1-\nu)$ and
$\pm\nu N$ at each cusp. For the split baryon, which extends between
the cusps at $\theta=0$ and $\pi$,
we find $\pm N$ since
the baryon must be a color singlet. However, we found new solutions, which
extend between the cusps at the same $\theta(=0$ or $\pi)$ as shown below.
In this case, for the solution with two cusps at $\theta=\pi$, we must choose the 
flux-combination as $\pm N(1-\nu)$ and $\mp N(1-\nu)$. And for the one
with the cusps at $\theta=0$, the flux should be assigned as
$\pm \nu N$ and $\mp \nu N$. Then the total flux is zero
in both cases, and we call these solutions as baryonium.}


In the case of $q>0$, $r'(\pi)$ or $r'(0)$ does not diverges any more, but 
these points, $\theta=\pi$ and $\theta=0$, are singular and tension 
to deform the D5 brane is observed. This tension can be cancelled out 
by adding the fundamental strings whose number is given by the bundles
observed for $q=0$. This picture is very naturally understood since the 
system of the brane and the fundamental strings deforms continuously
with the scale parameter $q$.


{Indeed above mentioned statements can be explisitly checked.  
For this purpose we calculate the tensions at the cusps 
\cite{GI, GINT, BLL} in Appendix B.  
To compare tensions, we take a ``vertical limit'', namely, $r'_1\to\infty$, 
$r^{(1)}_x\to\infty$. Then the following equality holds; 
\beq \label{no-force}
{{\delta U}\over{\delta r_1}}=(1-\nu)N{{\delta U_F}\over{\delta r_1}}\,.
\eeq
The tensions in the $x$-direction vanish in the vertical limit.  
The above  equality means the tension of the cusp equals to 
$(1-\nu)N$ times that of F-string automatically in the vertical limit 
in the case of $q>0$.}{ Similary, we find the $\nu N$ times tension of F-string
on the other cusp.}


\section{Baryonium States}\label{bound}

\vspace{.23cm}
{\bf Equations of motion} 
\vspace{.3cm}

In terms of (\ref{u}), we could obtain two kinds of baryon configurations
\cite{cgs,GI}. In both cases, we should notice that the  
solutions $r(\theta)$ and $x(\theta)$ cover
whole region of $\theta$, $0\leq\theta\leq\pi$. However there are other kinds of solutions, which
cover only a part of the variable $\theta$, i.e. 
(i) $\theta_0\leq \theta\leq \pi$ or (ii) $0\leq \theta\leq \theta_1$, 
where $\theta_0>\theta_1$.

\vspace{.3cm}
Here we identify these type of solutions embedded in (i) or (ii)
as the baryonium. 
Namely,
for the baryonium solution, $\theta$ does not cover all the region for both cases.
For example, in the case of (i), 
the solution $x(\theta)$ starts
at $x(\pi)(= x_-<0)$ and passes $x(\theta_0)\equiv 0$ smoothly, then
arrived at $x(\pi)(= x_+>0)$. So, this configuration extends from $x_-$ to
$x_+$ in our real three space. Here we should notice that the two end points are at
$\theta=\pi$ then the flux numbers at these points must be the same but with
opposite direction. This
corresponds exactly to the baryonium as mentioned in the introduction.
Similarly, we can consider the baryonium of region (ii), whose end points 
are at $\theta=0$ with different flux numbers.

\vspace{.3cm}
The difference of the baryon and baryonium solutions is reduced to the difference of their boundary conditions. 
Generally speaking, in solving the differential equations,
the boundary condition determines the integration constants which correspond to the constant of motion
like energy. In the present case also, we introduce such a constant, which
discriminates the solution of baryon and baryonium.


\vspace{.2cm}
In order to 
introduce such a constant, we rewrite (\ref{u}) by changing the integration variable
from $\theta$ to $x$ as follows
\be \label{upar}
U = {N\over 3\pi^2\alpha'}\int dx~e^{\Phi/2}
\sqrt{ r^2\dot{\theta}^2+\dot{r}^2+(r/R)^{4}}~
\sqrt{V_{\nu}(\theta)},
\ee
where dots denotes the derivative with respect to $x$. 
We can introduce an integral constant $h$ as
a ``Hamiltonian'' for the corresponding ``time variable'' $x$ as follows
\beq \label{h}
 h=\dot{r}p_{r}+\dot{\theta}p_{\theta}-L\, ,\quad
\eeq
where 
\beq
 L=e^{\Phi/2} \sqrt{ r^2\dot{\theta}^2+\dot{r}^2+(r/R)^{4}}~\sqrt{V_{\nu}(\theta)}
\eeq
\beq
 p_{r}={\partial L\over \partial \dot{r}}=\dot{r}Q\, , \quad
 p_{\theta}={\partial L\over \partial \dot{\theta}}=r^2\dot{\theta}
           Q\, ,
\eeq
and 
\beq \label{Q}
Q=\left({R\over r}\right)^2\sqrt{e^{\Phi}V_{\nu}-
   \left({p_{\theta}^2\over r^2}+p_{r}^2\right)}
\eeq
Then $h$ is written in terms of the momentum as
\beq\label{h2}
 h 
=-\left({r\over R}\right)^2\sqrt{e^{\Phi}V_{\nu}-
   \left({p_{\theta}^2\over r^2}+p_{r}^2\right)}\, ,
\eeq
and the equations of motion are obtained as
\beq\label{dot-r-th}
\dot{r}=
{p_r\over Q}\, , \quad
\dot{\theta}=
{p_{\theta}\over r^2Q}\, ,
\eeq
\beq\label{dot-pr-th}
 \dot{p}_r=-{\partial h\over \partial r}\, , \quad
 \dot{p}_{\theta}=-{\partial h\over \partial \theta}~.
\eeq
These equations are convenient to find the baryonium vertex solution as seen
below.

\begin{figure}[htbp]
\begin{center}
  \includegraphics[width=9cm]{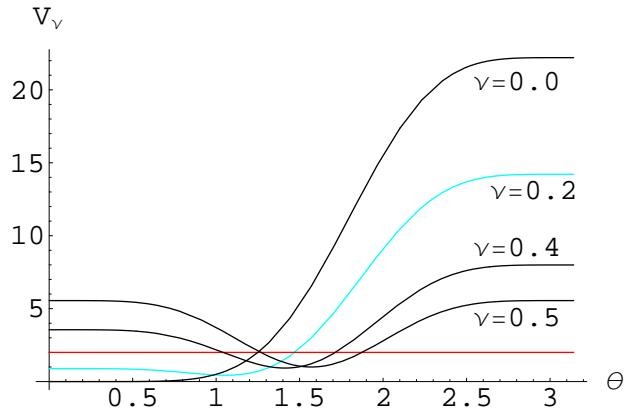}
\caption{\small $V_{\nu}(\theta)$
for $\nu=0.2,~0.4$, and $0.5$. The horizontal line shows a sample line
of $V_{\nu}(\theta_0)={R^4h^2\over r_0^4+q}=2.0$. The crossing points between this line and the curve of
$V_{\nu}$ represnets $\theta_0$.\label{Vnu}}
\end{center}
\end{figure}


\vspace{.3cm}
Before giving explicit solution, we show how the type of solutions
is controlled by $h$.
From the definition of $h$ given in Eq.~(\ref{h}), we obtain
\beq\label{consth2}
 {p_{\theta}^2\over r^2}+p_r^2=\left({R\over r}\right)^4
\left({r^4+q\over R^4}V_{\nu}-h^2\right)
\eeq
then from the reality of the solution, the next constraint is obtained
\beq\label{consth3}
{r^4+q\over R^4}V_{\nu}\geq h^2~.
\eeq
At the end points, $x=x_{\pm}$, $\theta=\pi$ and $r$ could become large, however $r$
takes its minimum $r\simeq r_0$ at the mid point $x=0$. Expressing as 
$\theta\vert_{x=0}=\theta_0$
at this point, the value of $\theta_0$ is obtained by solving the equation,
\beq\label{consth4}
{r_0^4+q\over R^4}V_{\nu,\theta_0}=h^2~.
\eeq
This equation has two solutions for $\theta_0$ when $h$ is large 
as shown in the Fig. \ref{Vnu}, but there is one or no solution
for small $h$. The solutions of the former (latter) case of large (small) $h$
are identified with the baryonium (baryon).

\vspace{.3cm}
Consider the baryonium solution.
From (\ref{consth4}), $r_0$ is given as 
\beq\label{r0}
 r_0=\left({R^4h^2\over V_{\nu}(\theta_0)}-q\right)^{1/4}\, .
\eeq

\vspace{.3cm}
While the exact solutions are obtained by solving the above four Hamiton equations
(\ref{dot-r-th}) and (\ref{dot-pr-th}) as given below, we show an approximate
solution in the region where we can assume $\dot{r}\sim 0$
and $r\sim r_0$ near $x=0$. Under this assumption, we obtain from (\ref{consth2})
\beq
 \dot{\theta}=\pm{r_0\over R^4h}\sqrt{r_0^4+q}\sqrt{V_{\nu}(\theta)-V_{\nu}(\theta_0)}
\eeq
then we solve this as
\beq
 x(\theta)=\pm\int_{\theta_0}^{\theta}d\theta~
   {R^4h\over r_0\sqrt{r_0^4+q}\sqrt{V_{\nu}(\theta)-V_{\nu}(\theta_0)}}
\eeq
This solution is symmetric with respect to $x=0$ axis in $x$-$\theta$ plane.
The important point of this approximate
solution is that the solution runs from $x=0$ 
to two opposite directions, however they are going to the same pole on $S^5$ 
but with different values of $x$. In order to see the behavior of the solution
far from $\theta_0$, we must solve the exact form of equations. Actually we can find
the solutions as stated above exactly, namely they satisfy this symmetry at all 
$\theta$ even if the assumption
imposed here is not satisfied.

\vspace{.3cm}
Which pole, $\theta=0$ or $\pi$, is chosen depends on the value of $\theta_0$.
Notice that $V_{\nu}(\theta)$ has a minimum
at $\theta_c$ which is given as a slolution of 
\be \label{nuc}
\pi\nu=\theta_{c}-\sin\theta_{c}\cos\theta_{c}~.
\ee
and we find the minimum value as $V_{\nu}(\theta_c)=\sin^6(\theta_c)$.
This implies that the pole $\theta=0$ ($\theta=\pi$) is chosen for 
$\theta_0<\theta_c$ ($\theta_0>\theta_c$).
This is understood well from the Fig. \ref{Vnu}.
We notice however that the number of $\theta_0$ depends on the value of $h$
as seen for $\nu=0.2$ as seen in the Fig. \ref{Vnu}.
The situation is however changed by the value of $h$ to find two $\theta_0$ also
for $\nu=0.2$. 

But for the 
case of $\nu=0$, there is only one $\theta_0$ for any value of $h$. 
In this case, the baryonium
is constructed by $N$-quarks and anti-$N$-quarks attached at each end points.
In other words, we obtain a bound state of baryon and anti-baryon without any
loss of quark and anti-quark by their pair annihilation.

On the other hand,
for the case of $\nu=1/N$, we expect another interesting baryonium configuration
which is constructed by one quark and one anti-quark. This state is very similar to
the usual mesons, but it is different from them in the point that
the D5 vertex is included except for a quark and an anti-quark pair
in this state. 

\vspace{.7cm}
{\bf Numerical solutions}

\vspace{.3cm}
Here we give the explicit baryonium solutions mentioned above. The equations
are solved numerically since it would be impossible to solve them analytically.
Firstly, we give a way to obtain the symmetric solutions. It is convenient to
set the following boundary conditions at $x=0$ as
\beq \label{baryoniumbc}
 \theta(0)=\theta_0\, , \quad r(0)=r_0\, , \quad {\rm and}~~
  p_r(0)=p_{\theta}(0)=0
\eeq
They are given as follows. First, an appropriate $\theta_0$ is given for fixed
$h$, then $r_0$ is determined by the above relation (\ref{r0}). 
The last two conditions, $p_r(0)=p_{\theta}(0)=0$, 
are necessary to obtain a baryonium solution which is symmetric with respect
to $x=0$ axis. 

\begin{figure}[htbp]
\vspace{.3cm}
\begin{center}
  \includegraphics[width=5.5cm,height=5cm]{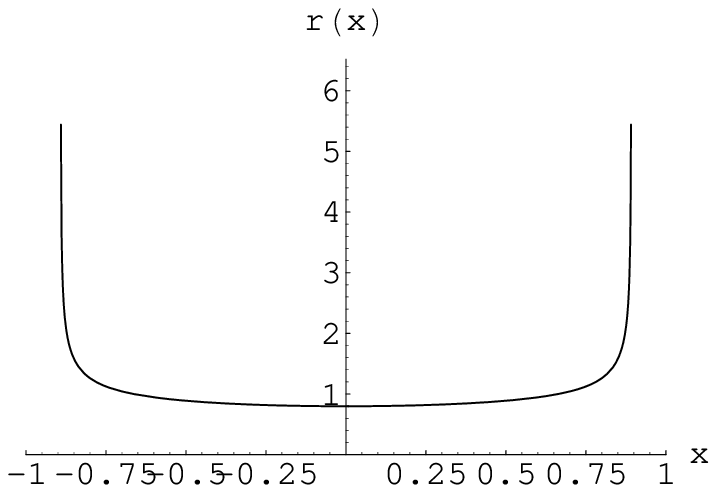}
  \includegraphics[width=5.5cm,height=5cm]{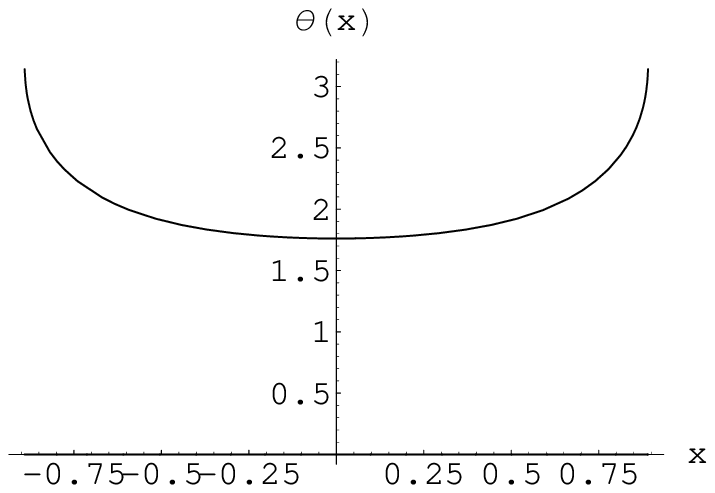}
\caption{\small The typical $D5/ \overline{D5}$ solution for $q=0.3$, $\nu=0.5$
at $h=-1$ and $\theta_0=1.6708$. The boundary conditions are $p_r(0)=p_{\theta}(0)=0$.\label{sol-1}}
\end{center}
\end{figure}

\begin{figure}[htbp]
\vspace{.3cm}
\begin{center}
  \includegraphics[width=7cm,height=5cm]{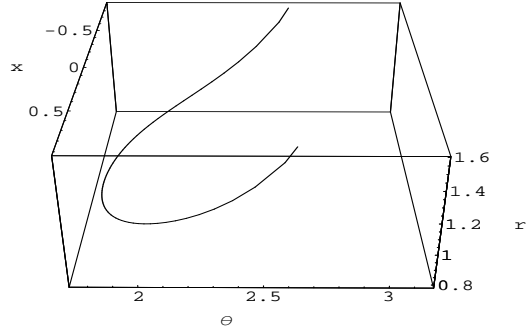}
\caption{\small The 3d graphic of $D5/ \overline{D5}$ solution 
given in the Fig. \ref{sol-1} for $q=0.3$, $\nu=0.5$
at $h=-1$ and $\theta_0=1.6708$.\label{sol-2}}
\end{center}
\end{figure}

\vspace{.3cm}

An explicit example of the baryonium solution is obtained
for $\nu=0.5$, $h=-1$ and $\theta_0=1.6708$ under the above boundary
conditions (\ref{baryoniumbc}). The results
are shown in the
Figs. \ref{sol-1} and \ref{sol-2}. From the Fig. \ref{sol-1}, we can
easily understand the fact that this solution is
interpreted as the $D5/\overline{D5}$ bound state solution.
This kind of solutions can be obtained at any $\nu$, and
it can be considered as the vertex part of
a baryonium of $(1-\nu)N$ quarks and $(1-\nu)N$ ant-quarks
which are attached at each end point of the vertex for general $\nu$. 

Here, in the Fig.~{\ref{sol-4}, we show the $\nu=0.7$ case as an example.  


\begin{figure}[htbp]
\vspace{.3cm}
\begin{center}
\includegraphics[width=5.5cm,height=5cm]{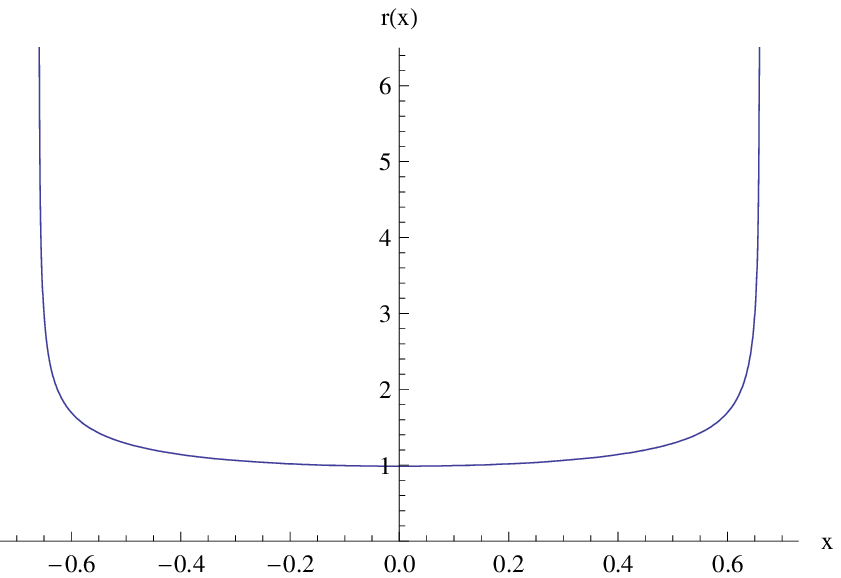}
\includegraphics[width=5.5cm,height=5cm]{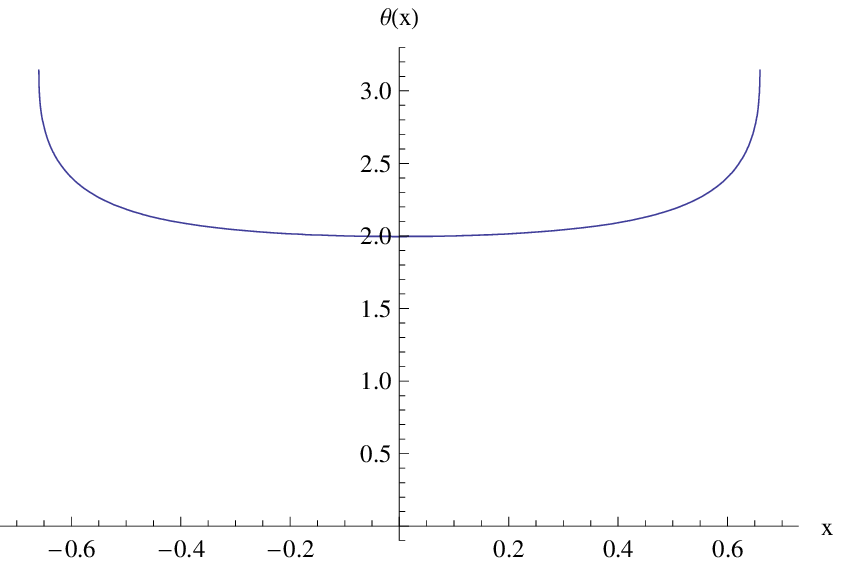}
\caption{$D5/\bar{D5}$ solution for $q=0.3$, $\nu=0.7$ at $ h=-1$ and
 $\theta_0=1.9962$. 
The boundary conditions are $p_r(0)=p_{\theta}(0)=0$.
\label{sol-4}}
\end{center}
\end{figure}

\vspace{.6cm}
\noindent{\bf Asymmetric solution}

\vspace{.3cm}
While the symmetric solutions are examined above,
the various asymmetric solutions are also obtained
when we solve the same equations with a slightly different boundary
conditions from the one of the symmetric solutions.
For example,
they are obtained by changing the boundary conditions, (\ref{baryoniumbc}). 
These solutions are also regarded as the baryonium since they connect
two cusps in the same side, at $\theta=\pi$ (or at 0) at different $x$. 

{
An example of a little asymmetric solution is shown in the Fig. \ref{sol-5}, 
where boundary conditions are changed to  $p_r(0)=0$ and  
$p_{\theta}(0)=10^{-4}$.

\begin{figure}[htbp]
\vspace{.3cm}
\begin{center}
  \includegraphics[width=5.5cm,height=5cm]{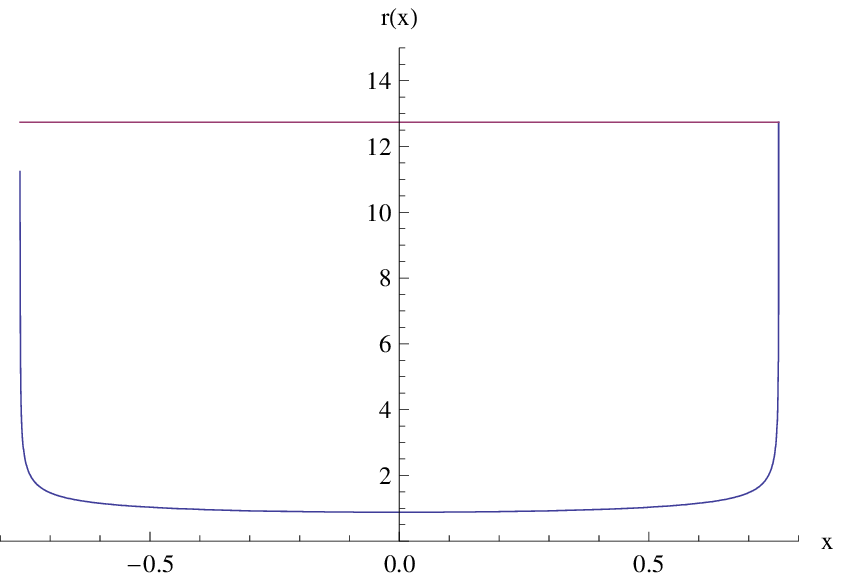}
  \includegraphics[width=5.5cm,height=5cm]{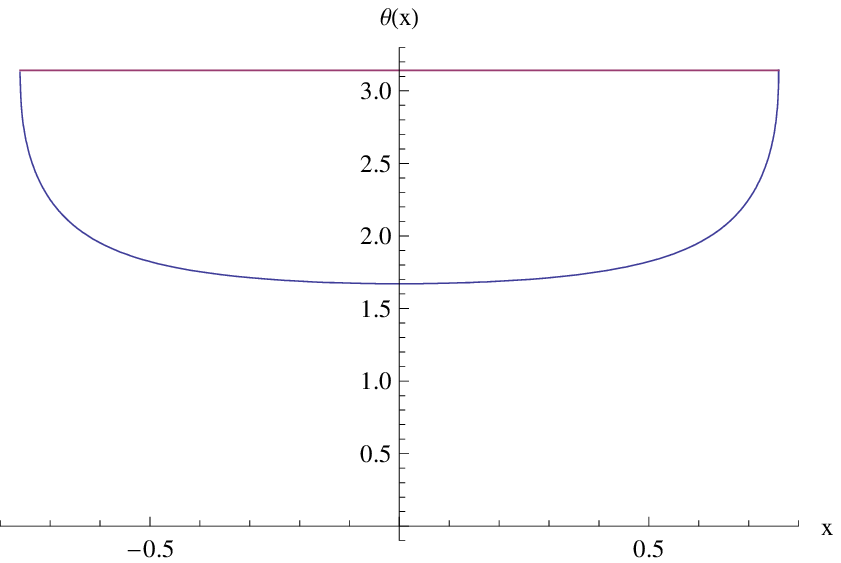}
\caption{\small The typical $D5/ \overline{D5}$ solution for $q=0.3$, $\nu=0.5$
at $h=-1$ and $\theta_0=1.6708$. The boundary conditions are $p_r(0)=0$, 
$p_{\theta}(0)=10^{-4}$.\label{sol-5}}
\end{center}
\end{figure}
}

In general, the energy of the asymmetric solution becomes higher
than the symmetric one. In this sense, the stable configurations can be
considered as the maximally symmetric one. Then, we consider hereafter 
symmetric solutions.

Another kind of solution obtained by a different boundary conditions is the
one called as the split baryon, which represents a baryon vertex. We give comments
to this solution in the next section.

\vspace{.5cm}
\noindent{\bf Stability of the baryonium vertex}

\vspace{.3cm}
In the next, we study the stability of the baryonium solution obtained above.
 From the viewpoint of energy, we concentrate on the symmetric vertex 
solutions, which are discriminated by its length $L$.
Here $L$ is defined as 
\beq
 L\equiv x_+(\pi)-x_-(\pi)\, , 
\eeq
where we assume $x_+(\pi)>x_-(\pi)$. Depending on the boundary
value $\theta_0$, 
both the $L$ and the vertex energy $U$, which is given by (\ref{upar}), vary. 
So, by varying $\theta_0$ for fixed $h$, the
relation between $U$ and $L$ is examined for the symmetric solutions. 
This relation could give us a critical
check for the stability of the baryonium configuration.

\begin{figure}[htbp]
\vspace{.3cm}
\begin{center}
  \includegraphics[width=7.5cm]{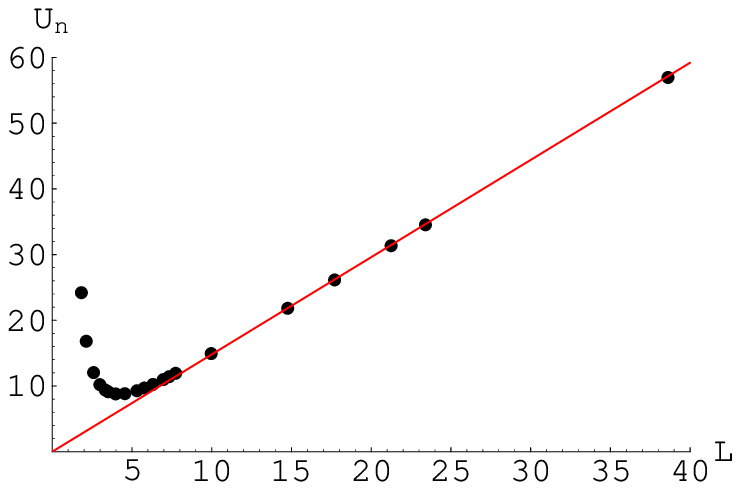}
  \includegraphics[width=7.5cm]{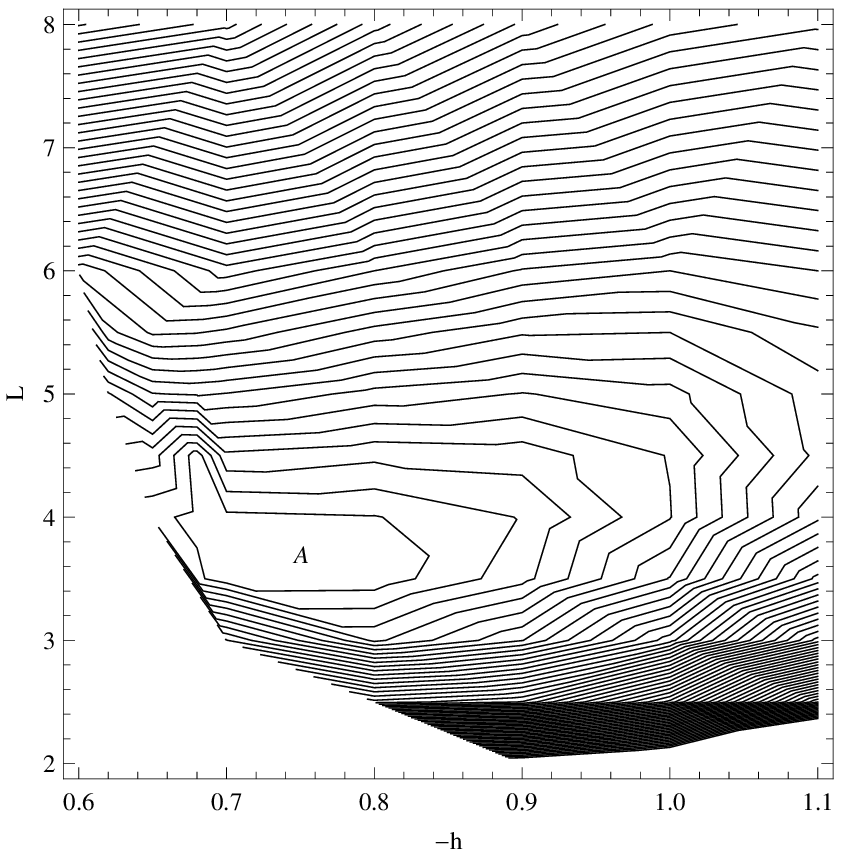}
\caption{\small The left: The normalized vetex energy $U_n=U/({N\over 3\pi^2\alpha'})$ versus its length in the x-direction
for $D5/\overline{D5}$ solution 
for $\nu=0.5$, $q=0.3$, $R=1$ and $h=-1$. The line represents $U_n=1.48 L$.
The right: The equi-potential curves in ($-h$,$L$) plane. The point A denotes
the bottom of the potential.
\label{U1-L05}}
\end{center}
\end{figure}

The numerical results of $U$ and $L$ for $h=-1$ and $\nu=0.5$ are shown 
in the left hand side of Fig. \ref{U1-L05}, where the normalized energy density, 
$U_n=U/({N\over 3\pi^2\alpha'})$, is shown instead of $U$. 
 From this
we can see that a clear minimum is found at definite and finite value
of $L$. Namely, the state for large $L$ and also for small $L$ needs large
energy to form it. Especially, the latter fact means that the baryonium vertex does not
vanish to $L=0$. 
For other values of $h$, similar $U$-$L$ relations are
expected and this is assured as follows. 

Here, $U$ depends on the two parameters
$h$ (or $r_0$) and $\theta_0$ except for the external parameters $R$ and $q$.
Then we can write as $U=U(h,r_0)$ or $U=U(h,\theta_0)$. Meanwhile 
it is possible to replace one of the parameters by a physical quantity,
for example $L$, as $U=U(h,L)$.

In order to assure the minimum of $U$ in the two dimensional parameter space, 
we show the equi-potential curves (contour for $U$)
in the $h$-$L$ plane. The numerical results are
shown in the right hand side of the Fig. \ref{U1-L05}. 
 From this, we find a minimum does exist near the point A, and we 
find that the minimum given in the left figure is near the same $L$ of point A.
In general, we can consider many kinds of paths in
this plane to study similar $U$-$L$ relation along it.

\vspace{.3cm}
 From the above results, we could read the following points.

\vspace{.3cm}
(i) The energy density $U$ has a minimum at a finite $L$. This implies the stability
of the baryonium configuration against the tachyon, which would appear
as strings connecting between D5 and anti-D5 branes. The mass square of this
string is given as $L^2/R^4-1/(2\pi\alpha')$ \cite{Sen}, then it is stable
for enough large $L$. Here the size of $L$ is measured by the scale $R$, the
AdS$_5$ radius, so we can set the parameters to satisfy the inequality,
\beq
  L^2/R^4>{1\over 2\pi\alpha'}\, ,
\eeq 
at least around the $L$ where $U$ is minimum.

\vspace{.3cm}
(ii)
Secondly, at large $L$, $U$ increase linearly with respect to $L$, and
we can approximate its behavior as
\beq
 U=\tau_B L\, .
\eeq
where $\tau_B$ denotes the tension of the baryonium vertex. The value in the case
given in the Fig. \ref{U1-L05} is evaluated and shown by a line.
However, we must notice that 
the above result is obtained for $h=-1$ and $\tau_B$ depends on $h$ as
seen from the right hand figure of $U(h,L)$. So we must be careful to study the
tension of the baryonium vertex. On this point, we do not discuss
furthermore.

\section{Baryonium and split baryon}
\label{baryonium}

As discussed in \cite{cgst,GI},
the baryon vertex has two types of solutions, that is, the point and split one.
The latter is similar to baryonium vertex in the sense that it has a finite 
length in our space.

As mentioned above, the baryon vertex configurations are also obtained
by solving the same equations with somewhat different boundary conditions. 
As given in \cite{GI,
GINT}, an easy way to obtain the baryon vertex, which extends also
in the $x$-direction with the two end points $x(\pi)$ and $x(0)$, is as 
follows.  First, set the boundary condition at $\theta=\theta_c$, where 
$\theta_c$ is the minimum point of $V_{\nu}(\theta)$ 
as given in (\ref{nuc}), as
\beq \label{pc}
 p_{\theta}\vert_{\theta=\theta_c}=\eta \neq 0~, \quad
 p_r\vert_{\theta=\theta_c}=0~.
\eeq
This boundary condition 
is necessary to embed the D5 brane on the whole region of $S^5$. In other
words, the polar angle $\theta$ must cover the whole range $0\leq \theta\leq\pi$,
and this becomes possible under the condition $\eta \neq 0$ in (\ref{pc}).
{This is the important point to identify the 
obtained configuration with
the baryon vertex. Namely, the two end points are at different polar angles,
$\theta=0$ and $\theta=\pi$.} 

\vspace{.3cm}
For the split baryon vertex obtained in this way, we examine its
energy. Roughly speaking, its configuration can be separated to the two parts of
extending to $x$ and to $r$ directions. And each part shears the energy of the
vertex. We find that the split baryon vertex can be smoothly deformed to the point
vertex by suppressing its energy. Then the configuration of minimum energy is
realized by the point vertex solution $r=q^{1/4}$ \cite{GI,GINT}.
This point is the important difference from the case of the baryonium, which
could not be pushed to a point in spite of the fact that
the baryonium is made of $D5/\overline{D5}$. 

As for the total mass of the baryon or baryonium, we must add F-strings
under the appropriate conditions called as no force condition \cite{GI}.
In this case, we could find that the minimum energy of split baryon is realized
when the length of the F-strings vanishes but the vertex length is finite.

Similar situation is also expected for the case of the baryonium. 
We examined the energy of baryonium 
for $\nu=0.5$, $q=0.3$, $R=1$, $r_{\rm max}=10$ and $h=-1$. 
Then we could find that the minimum of the energy is found when
the F-string length vanishes as in the split baryon.
However the details of the analyses are not given here.
We will give them in the near future.

\section{Summary and Discussion}\label{comparison}

We find a baryonium solution, which can be interpreted as bound state 
of baryon and anti-baryon, by solving the equations of motion for the
D5 brane action. The reason why the bound state $D5/\overline{D5}$ is  
obtained from the D5 brane action
is that the action used to solve the equations contains
the displacement flux operator $D$ in the squared form $D^2$. This fact enables
us to obtain the baryonium configuration, which is made by connecting
a D5 brane and its anti-brane, from a D5 brane action which we used. 

The configuration of the baryonium vertex looks like a string in our 
three dimensional space
and we find that its size or length $L$.
Its energy $U$ depends on the 
length $L$, and we can show the minimum of $U$ is realized at finite $L$. Then 
the size is kept finite in its stable state, and we could assure that its
configuration in the bulk is in a form that the $D5$
and $\overline{ D5}$ are separated enough not to be destabilized by the tachyon.
Then we can say that the baryonium state found here would be stable and 
it would be difficult to observe its decay into mesons. This would
be related to the selection rule for hadronic decay and the 
resultant narrow width of the baryonium\cite{FWR}.

The baryonium vertex solution given here is similar to a baryon vertex configuration
which also looks like a string in our three dimensional space. And they both
are obtained from the same equations of motion. However, one of the values of 
the displacement $D$ at the two cusps for the baryonium is different from the baryon.
The sum of $D$ at the two cusps is zero for the baryonium, but it is $N$ 
for the baryon. The energy minimum for the baryon is realized for $L=0$, which
looks like a point in our space. Meanwhile, its configuration looks string like
in the bulk.

A real baryonium should be made of the vertex and fundamental strings
attached at cusps as performed in the case of baryons. 
Then the mass spectra for the baryonium are examined to compare the spectra
given in recent experiments\cite{Belle}. The tetra-quark meson corresponds to the baryonium of $\nu=1/3$ and $N=3$. We can estimate the mass spectrum of this state. This would be given in the near future.

\section*{Appendix A}
Here we show another formulation of solving the equations of motion
derived from (\ref{u}) used in \cite{cgst,GI}.
Firstly, rewrite (\ref{u}) in terms of a 
general
worldvolume parameter $s$ defined by
functions $\theta=\theta(s)$, $r=r(s)$, $x=x(s)$ as:
\be \label{upar2}
U = {N\over 3\pi^2\alpha'}\int ds~e^{\Phi/2}
\sqrt{ r^2\dot{\theta}^2 + \dot{r}^2+(r/R)^{4}\dot{x}^2}~
\sqrt{V_{\nu}(\theta)},
\ee
where dots denote derivatives with respect to $s$.
Then the momenta conjugate to $r$, $\theta$ and $x$ are given as
\be \label{mom}
p_r=\dot{r}\Delta, \quad
p_{\theta}=r^2\dot{\theta}\Delta, \quad
p_{x}=(r/R)^{4}\dot{x}\Delta, \quad
\Delta=e^{\Phi/2}\frac{\sqrt{V_{\nu}(\theta)}}
   {\sqrt{ r^2\dot{\theta}^2 + \dot{r}^2+(r/R)^{4}\dot{x}^2}}~.
\ee
Since the Hamiltonian that follows from the action\req{upar} vanishes 
identically
due to reparametrization invariance in $s$. 
Then we consider the following identity
\be \label{ham}
2\tilde{H} =p_r^2 + \frac{p_{\theta}^2 }{r^2}+\frac{R^4}{r^4}p_{x}^2-
\left(V_{\nu}(\theta) \right) e^{\Phi}=0~.
\ee
Regarding this constraint as a new Hamiltonian, we obtain
the following canonical equations of motion, 
\bea\label{Heom}
\dot r &=& p_r~,~~ \dot p_r =\frac{2}{r^5}p_x^2 R^4
     + \frac{p_\theta^2}{r^3}+{1\over 2}
        \left(V_{\nu}(\theta)\right)~ e^{\Phi}\partial_r\Phi,\\
\dot\theta &=&\frac{p_\theta}{r^2}~, ~~\dot p_\theta = -6\sin^4\theta \left(\pi\nu-\theta+ 
\sin\theta\cos\theta
\right)~e^{\Phi}, \\
\dot x &=& \frac{R^{4}}{r^{4}}p_{x},~~\dot{p}_{x}= 0
\eea
The initial conditions should be
chosen such that $\tilde H=0$. By solving these equations, we could find
the same solutions given above.


\section*{Appendix B}

Here we calculate the tensions at the cusps \cite{GI, GINT, BLL}.  
In the present 
model the tension has $r$-component and $x$-component generally.  
Denoting $r_1=r(\pi)$ and $x_{1\pm}=x_{\pm}(\pi)$, where 
$x_{1-}<0$, and $x_{1+}>0$, the tensions are given by, 
\beq \label{d5tension1}
{{\delta U}\over{\delta r_1}}={(1-\nu)N\over{2\pi\alpha'}}{{e^{\Phi(r_1)/2}r'_1}\over{\sqrt{r_1^2+r_1'^2+(r_1/R)^4x_{1\pm}'^2}}},\qquad
{{\delta U}\over{\delta x_{1\pm}}}={(1-\nu)N\over{2\pi\alpha'}}{{e^{\Phi(r_1)/2}(r_1/R)^4x'_{1\pm}}\over{\sqrt{r_1^2+r_1'^2+(r_1/R)^4x_{1\pm}'^2}}}\,.
\eeq
In the above equation  
The factor $(1-\nu)$ comes from $|D(\nu,\theta=\pi)|$.  
  
On the other hand, tension of F-string is derived from the following action,
\beq \label{action-F}
U_F={1\over{2\pi\alpha'}}\int_{x_{1\pm}}^{x_{\rm max}}dx\,e^{\Phi/2}\sqrt{r_x^2+(r/R)^4}\,.
\eeq
where $r_x=\partial_r x$.  
Then the tension of the F-string is obtained as,
\beq \label{Ftension}
{{\delta U_F}\over{\delta r_1}}={1\over{2\pi\alpha'}}
{{e^{\Phi(r_1)/2}r_x^{(1)}}\over{\sqrt{r_x^{(1)2}+(r_1/R)^4}}}\,,
\eeq

To compare tensions, we take a ``vertical limit'', namely, $r'_1\to\infty$, 
$r^{(1)}_x\to\infty$. Then the following equality holds; 
\beq \label{no-force-2}
{{\delta U}\over{\delta r_1}}=(1-\nu)N{{\delta U_F}\over{\delta r_1}}\,.
\eeq
The tensions in the $x$-direction vanish in the vertical limit.  
The above  equality means the tension of the cusp equals to 
$(1-\nu)N$ times that of F-string automatically in the vertical limit 
in the case of $q>0$.  

Other case except vertical limit, the following no-force condtion,
\beq \label{no-force-rx}
{{\delta U}\over{\delta r_1}}=(1-\nu)N{{\delta U_F}\over{\delta r_1}}\,,\qquad
{{\delta U}\over{\delta x_{1\pm}}}=(1-\nu)N{{\delta U_F}\over{\delta x_{1\pm}}}\,,
\eeq
assures the relation between tensions of cusps and F-strings.


\section*{Acknowledgements}
This work was supported by the Grants from Electronics Research Laboratory, 
Fukuoka Institute of Technology, and M. Ishihara is also supported by 
JSPS Grant-in-Aid for Scientific Research No. 20 $\cdot$ 04335.

\vspace{.5cm}

\end{document}